\documentstyle[aps,prl,epsf]{revtex}

\def\be{\begin{equation}}
\def\ee{\end{equation}}
\def\bea{\begin{eqnarray}}
\def\eea{\end{eqnarray}}
\def\sp{{\mathcal C}}
\def\tr{{\rm tr}}
\def\1/2{\frac{1}{2}}
\def\¦{|}
\def\opone{\leavevmode\hbox{\small1\kern-3.8pt\normalsize1}}

\newcommand{\la}{\langle}
\newcommand{\ra}{\rangle}

\begin{document}

\draft

\title{Quantum non-locality in two three-level systems}

\author{A. Ac\'\i n$^{1,2}$, T. Durt$^3$, N. Gisin$^1$ and J. I. Latorre$^2$}

\address{$^1$ GAP-Optique, 20 rue de l'\'Ecole-de-M\'edecine, CH-1211 Geneva 4, Switzerland.\\
$^2$ Departament d'Estructura i Constituents de la Mat\`eria, Diagonal 647, E-08028 Barcelona, Spain.\\
$^3$ TONA, Vrije Universiteit Brussel, Pleinlaan 2, B-1050
Brussels, Belgium.}
\date{\today}

\maketitle


\begin{abstract}
Recently a new Bell inequality has been introduced
\cite{CGLMP,KKCZO} that is strongly resistant to noise for
maximally entangled states of two $d$-dimensional quantum systems.
We prove that a larger violation, or equivalently a stronger
resistance to noise, is found for a non-maximally entangled state.
It is shown that the resistance to noise is not a good measure of
non-locality and we introduce some other possible measures. The
non-maximally entangled state turns out to be more robust also for
these alternative measures. From these results it follows that two
Von Neumann measurements per party may be not optimal for
detecting non-locality. For $d=3,4$, we point out some connections
between this inequality and distillability. Indeed, we demonstrate
that any state violating it, with the optimal Von Neumann
settings, is distillable.
\end{abstract}

\pacs{PACS Nos. 03.67.-a, 03.65.Bz}

\section{Introduction}

Since the seminal work of Bell \cite{bell}, it is known that no
local variable (LV) theory \cite{EPR} can reproduce all the
statistical results predicted by Quantum Mechanics for states of
composite systems. In fact, it was proven that the correlations
observed between two spin-$\1/2$ particles in the singlet state
violate some inequalities, called Bell inequalities, that any LV
theory satisfies. This provided a possible definition of quantum
non-locality: a quantum state is said to be non-local when it
violates a Bell inequality. More recently, it was shown that any
pure state that is not separable \cite{sep}, i.e. such that the
parties cannot prepare it using only local operations and
classical communication (LOCC), violates a Bell inequality
\cite{Gisin,PR}. Unfortunately, the distinction between local and
non-local states is not as clear for density matrices. Indeed
there exist mixed states that, despite being entangled, do not
violate any Bell inequality \cite{Werner} (see however
\cite{Popescu}). Consequently, there are two different ways of
considering quantum non-locality: entanglement (quantum non-local
resources are required for preparing the state) and violation of
Bell inequalities.

The interest in entanglement has dramatically increased during the
last two decades due to the fact that entanglement is the key
ingredient in many of the recent quantum information applications.
Many efforts have been devoted to quantify entanglement (see for
instance \cite{BBPS,entquant}) as a resource, and nowadays
bipartite pure-state entanglement is well understood. The
maximally entangled state of a bipartite system,
$\¦\Psi\ra\in\sp^d\otimes\sp^d$, reads
\begin{equation}
\label{maxent} |\Psi\ra=\frac{1}{\sqrt d}\sum_{i=0}^{d-1}\¦jj\ra ,
\end{equation}
where $\¦j\ra$ are orthonormal bases in each subsystem. Given an
entangled state in $\sp^d\otimes\sp^d$, it is important to know if
it can be distilled, i.e. if $N$ copies of it can be transformed
by LOCC into $M$ copies of $\¦\Psi\ra$. State distillability, or
useful quantum correlations, offers an alternative way of
analyzing quantum non-locality. All bipartite entangled pure
states can be reversibly transformed using LOCC into $\¦\Psi\ra$
(in the so-called asymptotic regime). The ratio of the conversion
is equal to the entropy of entanglement \cite{BBPS}. For the
mixed-state case, however, the picture is again not clear: indeed
it is not known when a given entangled density matrix is
distillable. It is also not known how distillability properties of
mixed states are connected to Bell inequality violation
\cite{BIdist,Terhal,KZG,AS}.

All these questions about non-locality in mixed states are
essentially solved for the simplest case of two two-level systems,
also called {\sl qubits}. There, the Peres-Horodecki criterion of
positivity of the partial transposition \cite{Peres,Horo} detects
if a given state is separable or entangled. Furthermore all
two-qubit entangled states are distillable \cite{Horo2}. As far as
Bell inequalities is concerned, the CHSH-inequality \cite{CHSH}
plays a very important role \cite{Fine}, and it is already known
when a quantum state violates it \cite{Horo3}. Its maximal
violation is only obtained for the maximally entangled state of
two spin-$\frac{1}{2}$ particles.

Recently an inequality \cite{CGLMP,KKCZO} has been found that
generalizes the CHSH inequality to systems of arbitrary dimension,
$d$, often referred as {\sl qudits}. This offers the opportunity
of testing some of the concepts seen above for this inequality.
This is the scope of the present article. We see that,
surprisingly, the maximal violation of the inequality, under Von
Neumann measurements, is not obtained for the maximally entangled
state of two three-level systems or {\sl qutrits} (section III).
This leads us to analyze if the resistance to noise is a correct
measure of non-locality. By a simple example we see that it is
not. Furthermore, we prove that any state violating the inequality
with the optimal settings is distillable, and the witness that
comes from it \cite{Terhal} is decomposable (section IV).

\section{Bell inequality for qutrits}

In this section we review the Bell inequality for qutrits obtained
in \cite{CGLMP,KKCZO}. The two parties, $A$ and $B$, are allowed
to perform two different three-outcome measurements, $A_1$ and
$A_2$ for $A$, and $B_1$ and $B_2$ for $B$. Denoting by
$P(A_i=B_j+k)$ the probability that the outcomes for party $A$ and
$B$, measuring $A_i$ and $B_j$, differ by $k$ modulo $d$ (in this
case $d=3$), one can consider the following Bell inequality,
\begin{eqnarray}
\label{tritineq}
  I_3&=&P(A_1=B_1)+P(B_1=A_2+1)+P(A_2=B_2)+P(B_2=A_1)  \nonumber\\
  &-&P(A_1=B_1-1)-P(B_1=A_2)-P(A_2=B_2-1)-P(B_2=A_1-1)\leq 2 .
\end{eqnarray}

The authors of \cite{CGLMP,KKCZO} analyzed the violation of this
inequality by the maximally entangled state (\ref{maxent}) of two
qutrits. They consider the following settings: first the two
parties apply a unitary operation on each subsystem with only
non-zero terms in the diagonal equal to $e^{i\phi_a(j)}$ for $A$
and $e^{i\varphi_b(j)}$ for $B$, with $j=0,1,2$ and $a,b=1,2$.
These unitary operations are denoted by $U(\vec\phi_a)$, where
$\vec\phi_a\equiv(\phi_a(0),\phi_a(1),\phi_a(2))$, for party $A$,
and the same for $B$ with $\varphi$ instead of $\phi$. The values
of these phases are
\begin{equation}
\label{phase}
\phi_1(j)=0 \quad\quad \phi_2(j)=\frac{\pi}{3}j
\quad\quad \varphi_1(j)=\frac{\pi}{6}j \quad\quad
\varphi_2(j)=-\frac{\pi}{6}j ,
\end{equation}
with $j=0,1,2$. In this scenario, the freedom in the choice of the
measurement the parties apply is given by this first unitary
transformation. Then, party $A$ carries out a discrete Fourier
transform, $U_{FT}$ \cite{fourier}, and $B$ applies $U_{FT}^*$,
and finally they measure in the initial basis $\¦j\ra$. With these
experimental settings, Quantum Mechanics predicts that
$I_3(\¦\Psi\ra)=4(2\sqrt 3+3)/9\simeq 2.8729$. Numerical
simulations \cite{CGLMP,Zukowski} show that this is the maximum
value of $I_3$ achieved starting from the maximally entangled
state, $\¦\Psi\ra$. It is then conjectured that the described
experimental settings are optimal for $\¦\Psi\ra$ with this
inequality.

It is possible to define a more absolute measure of non-locality
in the following way: the initial entangled state, $\¦\Psi\ra$, is
mixed with some amount of noise, the resulting state being equal
to
\begin{equation}
\label{menoise}
  \rho=\lambda\¦\Psi\ra\la\Psi\¦+(1-\lambda)\frac{\opone}{9} ,
\end{equation}
when $0\leq\lambda\leq 1$. The entanglement in $\rho$ decreases
with $\lambda$, so one can look for the maximal amount of noise,
or minimum $\lambda$, such that it is still not possible to build
a LV model for the predicted probabilities. This measure of
non-locality, known as resistance to noise, depends on the
experimental settings, that is on the number and the type of
measurements each party can apply. The inequality (\ref{tritineq})
reproduces for (\ref{maxent}) the same resistance to noise as it
was found numerically in \cite{Zukowski}, with two Von Neumann
measurements on each side. This means that the probabilities
resulting from performing these measurements on the state
(\ref{menoise}) admit a LV model when
\begin{equation}
\label{meres}
  0\leq\lambda\leq\frac{2}{I_3(\¦\Psi\ra)}\simeq 0.6962 .
\end{equation}

\section{Maximal violation of the inequality}

The authors of \cite{CGLMP,KKCZO} focused their attention into the
violation of this inequality for the maximally entangled state of
two qutrits. However, it may happen that a larger value of $I_3$
is found if we consider a different initial state.

For the experimental settings (\ref{phase}), one can derive the
Bell operator \cite{BMR} associated to this inequality. The joint
probability, $P(A_a=j,B_b=k)$, of detecting result $j$ in $A$, $k$
in $B$, when $A_a$ and $B_b$ are measured and the initial state is
$\¦\Phi\ra\in\sp^3\otimes\sp^3$, is given by
\begin{equation}
  \tr\left(\Pi_j\otimes\Pi_k\,V(\vec\phi_a)\otimes
  V(\vec\varphi_b)\¦\Phi\ra\la\Phi\¦V(\vec\phi_a)^\dagger\otimes
  V(\vec\varphi_b)^\dagger\right)=\tr\left(V(\vec\phi_a)^\dagger\otimes
  V(\vec\varphi_b)^\dagger\Pi_j\otimes\Pi_k\,V(\vec\phi_a)\otimes
  V(\vec\varphi_b)\¦\Phi\ra\la\Phi\¦\right) ,
\end{equation}
where $V(\vec\phi_a)\equiv U_{FT}\,U(\vec\phi_a)$ and
$V(\vec\varphi_b)\equiv U^*_{FT}\,U(\vec\varphi_b)$. From this
formula the Bell operator, $B$, such that
\begin{equation}
\label{bop}
  I_3(\¦\Phi\ra)=\tr(B\¦\Phi\ra\la\Phi\¦)=\la B\ra_\Phi ,
\end{equation}
is found, and it reads
\begin{equation}
\label{bellop}
  B=\left(\matrix{0&0&0&0& \frac{2}{\sqrt 3} &0&0&0& 2 \cr
    0&0&0&0&0& \frac{2}{\sqrt 3} &0&0&0\cr
    0&0&0&0&0&0&0&0&0\cr
    0&0&0&0&0&0&0& \frac{2}{\sqrt 3} &0\cr
    \frac{2}{\sqrt 3} &0&0&0&0&0&0&0& \frac{2}{\sqrt 3}\cr
    0& \frac{2}{\sqrt 3} &0&0&0&0&0&0&0\cr
    0&0&0&0&0&0&0&0&0\cr
    0&0&0& \frac{2}{\sqrt 3} &0&0&0&0&0\cr
    2 &0&0&0& \frac{2}{\sqrt 3} &0&0&0&0\cr}\right) .
\end{equation}
The maximal violation of the inequality with these settings
corresponds to the maximum eigenvalue of $B$, which is equal to
$1+\sqrt{11/3}\simeq 2.9149$. Note that this value is a bit larger
than the violation obtained for $\¦\Psi\ra$. Indeed its
corresponding eigenvector is a non-maximally entangled state of
two qutrits that reads,
\begin{equation}
\label{maxv}
  \¦\Psi_{mv}\ra=\frac{1}{\sqrt n}\left(\¦00\ra+\gamma\¦11\ra+\¦22\ra\right) ,
\end{equation}
where $\gamma=(\sqrt 11-\sqrt 3)/2\simeq 0.7923$, and
$n=2+\gamma^2$ (the same results are obtained starting from the
inequality in \cite{KKCZO}). All the details of the calculation
are given in the appendix.

It is natural to ask about the optimality of the chosen set of
measurements. We have performed a numerical search for this
inequality, varying freely the two Von Neumann measurements
performed by each of the parties and the initial state. The
maximal violation is indeed obtained by the configuration shown
above. Moreover in the appendix we prove that these experimental
settings give a local maximum for the largest eigenvalue of $B$.

The Bell inequality (\ref{tritineq}) was also extended to
arbitrary dimension in \cite{CGLMP}. There it was shown that the
combination of joint probabilities
\begin{eqnarray}
\label{genineq}
  I_d=&&\sum_{k=0}^{[d/2]-1}\left(1-\frac{2k}{d-1}\right)
  \Big(P(A_1=B_1+k)+P(B_1=A_2+k+1)+
  P(A_2=B_2+k)+P(B_2=A_1+k) \nonumber\\
  && -P(A_1=B_1-k-1)-P(B_1=A_2-k)
  -P(A_2=B_2-k-1)-P(B_2=A_1-k-1)\Big)\leq 2 ,
\end{eqnarray}
for LV models. However this inequality can be violated if we
consider the maximally entangled state (\ref{maxent}) and similar
experimental settings with $\phi_1(j)=0$, $\phi_2(j)=j\pi/d$,
$\varphi_1(j)=j\pi/(2d)$ and $\varphi_2(j)=-j\pi/(2d)$, and
$j=0,\ldots,d-1$. Indeed, this inequality reproduces the
resistance to noise obtained numerically in \cite{Zukowski}, but
now for two Bell multiports \cite{tritter} on each side. Starting
from (\ref{genineq}) we can derive the corresponding Bell operator
and a larger violation is again found for partially entangled
states of two qudits. Table \ref{summary} summarizes these results
up to $d=8$. Note that the difference between the violation for
$\¦\Psi\ra$ and $\¦\Psi_{mv}\ra$ increases with the dimension.

These results are quite surprising. Previous numerical work in
\cite{Zukowski} showed that the resistance to noise for the
maximally entangled state of two qutrits with two Von Neumann
measurements per party is indeed the one predicted by this Bell
inequality (\ref{meres}). Nevertheless, there exists a
non-maximally entangled state, $\¦\Psi_{mv}\ra$, whose quantum
correlations are more resistant to noise, since its violation of
(\ref{tritineq}) is larger. Let us mention here that the parties,
if they start with the maximally entangled state $\¦\Psi\ra$, are
always able to prepare by LOCC and with probability one the state
$\¦\Psi_{mv}\ra$, and then to check the violation of the Bell
inequality. This leads us to analyze more precisely whether the
resistance to noise is a good measure of non-locality (see also
\cite{CP}).

Take the following two-qudit state,
\begin{equation}
\label{schmidt2}
  \¦\Psi_2\ra=\frac{1}{\sqrt 2}(\¦00\ra+\¦11\ra) ,
\end{equation}
where $d-2$ of the Schmidt coefficients are zero. Now, consider
the CHSH inequality \cite{CHSH}
\begin{equation}
\label{CHSHin}
  \langle\, A_1(B_1+B_2)+A_2(B_1-B_2)\,\rangle\leq 2 ,
\end{equation}
where $A_i$ and $B_i$, $i=1,2$, are measurements of two outcomes,
labelled by +1 and -1. The maximum violation of this inequality
attained by quantum states is $2\sqrt 2$ \cite{Cirelson}. The
following choice of measurements achieves this maximum for the
state (\ref{schmidt2}), \be A_1^{+1}=P_0 \quad\quad
A_2^{+1}=P_{\frac{\pi}{2}} \quad\quad B_1^{+1}=P_{-\frac{\pi}{4}}
\quad\quad B_2^{+1}=P_{\frac{\pi}{4}} , \ee where $P_{\omega}$ is
the projector onto the state $1/\sqrt
2(|0\rangle+e^{i\omega}|1\rangle)$, and
$X_i^{-1}=\opone-X_i^{+1}$, with $X=A,B$ and $i=1,2$. Since
$\tr(X_i^{+1})=1$ and $\tr(X_i^{-1})=d-1$, the contribution of the
maximally mixed noise to (\ref{CHSHin}) for these settings is not
zero. Indeed it is not difficult to see that in this case the
resistance to noise for $|\Psi_2\rangle$ is
\begin{equation}\label{wrongres}
  \lambda=\frac{1-\left(\frac{d-2}{d}\right)^2}
  {\sqrt 2-\left(\frac{d-2}{d}\right)^2} ,
\end{equation}
which tends to zero when $d\rightarrow\infty$! One can argue that
(\ref{schmidt2}) is not really a two-qudit state, but similar
results can be obtained for states of full Schmidt number that are
infinitesimally close to it. This example shows that the
resistance to noise is not a good measure of non-locality. Let us
briefly explore here other alternative candidates.

The first possibility consists on studying the resistance of
entangled states when they are mixed with the state resulting from
the tensor product of the reduced density matrices. Then, given an
entangled state $|\Phi\rangle\in\sp^d\otimes\sp^d$, we want to
determine the minimum value of $\lambda'$ such that there is no LV
model for the state
\begin{equation}\label{res1}
  \rho'=\lambda'|\Phi\rangle\langle\Phi|+
  (1-\lambda')\rho_A\otimes\rho_B ,
\end{equation}
where $\rho_A\equiv\tr_B(|\Phi\rangle\langle\Phi|)$ and similarly
for $B$. This measure has the advantage of being equal to the
resistance to noise for maximally entangled states, and avoids
problems as the one previously discussed. A second possibility can
be to consider mixtures of the initial entangled states with the
closest separable one, $\sigma_{AB}$. Similarly as for the
Relative Entropy of Entanglement \cite{relent}, one can choose the
relative entropy as a measure of distance,
$S(\rho,\sigma)=\tr(\rho\log\rho-\rho\log\sigma)$. Therefore
$\sigma_{AB}$ is defined as the state minimizing
$S(|\Phi\rangle\langle\Phi|,\sigma)$ over the set of separable
states \cite{note}. Now, we look for the minimum $\lambda''$ such
that the state
\begin{equation}\label{res2}
  \rho''=\lambda''|\Phi\rangle\langle\Phi|+
  (1-\lambda'')\rho_{AB}
\end{equation}
does not admit a LV description. Remarkably, for the states
(\ref{maxent}) and (\ref{maxv}), the settings defined above and
the inequality (\ref{tritineq}), the three numbers for each state
coincide, i.e. $\lambda_{min}=\lambda'_{min}=\lambda''_{min}$.
Thus, no change is observed by using these alternative measures of
non-locality.

All these reasonings suggest that two Von Neumann measurements are
not optimal for detecting non-locality in two-qutrit systems. A
possible way-out can be that more general measurements (POVM, that
is positive operator valued measures) are required for having a
larger violation for $\¦\Psi\ra$. It seems more likely that more
observables for each party, and a new Bell inequality, are needed.
Another interesting scenario consists on the analysis of these
quantum correlations under sequences of measurements as in
\cite{CP}. Indeed, states (\ref{menoise}) are entangled and
distillable for $1/(N+1)<\lambda\leq 1$ \cite{Horo4}. Finally, it
also follows from this result that it is not correct even for pure
states to quantify entanglement by means of the violation of a
particular Bell inequality \cite{MNW}.

\section{Distillability and Bell inequalities}

Violation of Bell inequalities is a possible manifestation of
non-locality, but, as it has been discussed in the introduction,
there are other ways of thinking about non-locality. From the
point of view of quantum information it is interesting to know if
the correlations in a quantum state are useful, i.e. if the state
is distillable. It is usually conjectured that Bell inequality
violation implies the distillability of the state \cite{BIdist}.
In this section we show that, for the experimental settings seen
above, the corresponding entanglement witness is decomposable.
This implies that any bound entangled state with positive partial
transposition \cite{bound} does not violate this inequality.
Moreover, from our construction, it can be proven that any state
violating the inequality is distillable.

From (\ref{tritineq}) it is possible to construct the Bell
operator \cite{BMR}, $B$, and from it the entanglement witness
\cite{Terhal}, $W=2-B$, such that $\tr(\rho_SW)\geq 0$ for all
separable states $\rho_S$, and there exists an entangled state,
$\rho$, that is detected, i.e. $\tr(\rho W)<0$. There exists a
class of entanglement witnesses, called decomposable, $W_d$, that
can be written as
\begin{equation}
\label{decomp}
  W_d=P+Q^{T_A} ,
\end{equation}
where $P$ and $Q$ are positive operators, and $T_A$ denotes
partial transposition with respect to subsystem $A$. Note that
these entanglement witnesses are not able to detect entangled
states with positive partial transposition, since if
$\rho^{T_A}\geq 0$ we have $\tr(\rho W)=\tr(\rho
P)+\tr(\rho^{T_A}Q)\geq 0$. Thus, they are not very useful for
checking the separability of a given state (see for instance
\cite{Horo,decw}), since they do not provide more information than
the partial transposition operation. In the following lines we
show that the witness coming from (\ref{tritineq}) is
decomposable.

\begin{center}
\begin{figure}
 \epsfysize=8 cm
 \epsffile{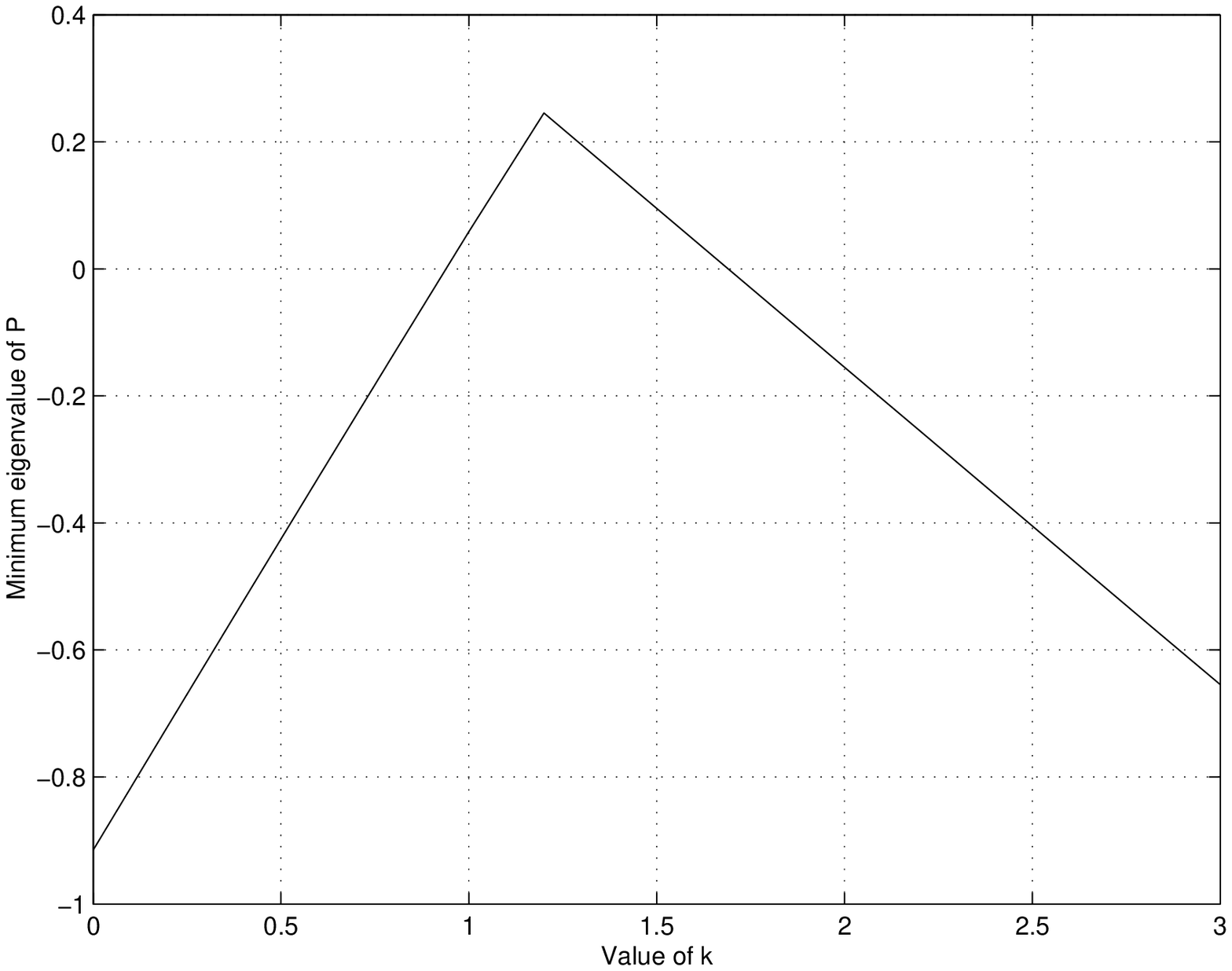}
\medskip
\caption{ Variation of the minimum eigenvalue of $P$ in
(\ref{decas}) with $k$.}
\label{firstdec}
\end{figure}
\end{center}

Our aim is to prove that there exist some positive operators, $P$
and $Q$ satisfying (\ref{decomp}). Note that the role of $Q$ is to
detect the entangled state, so we will choose an operator $Q$
maximizing
$\tr(\¦\Psi_{mv}\ra\la\Psi_{mv}\¦Q^{T_A})=\tr(\¦\Psi_{mv}\ra\la\Psi_{mv}\¦^{T_A}Q)$.
We will take then $Q$ proportional to the projector onto the space
of negative eigenvalues of $\¦\Psi_{mv}\ra\la\Psi_{mv}\¦^{T_A}$,
the state that gives the maximal violation of the inequality. If
we work in the Schmidt basis of this state, it is easy to see that
$Q$ is proportional to the projector onto the antisymmetric space
of two qutrits,
\begin{equation}
P_{a}=\1/2\sum_j\sum_{k\neq j}(\¦jk\ra-\¦kj\ra)(\la jk\¦-\la
kj\¦).
\end{equation}
Our guess then for the decomposition (\ref{decomp}) of $W$ is
\begin{equation}
\label{decas}
  W=P+kP_{a}^{T_A} ,
\end{equation}
where $k$ is a positive number to be determined. Now, we look for
a value of this constant such that $P=W-kP_{a}^{T_A}\geq 0$, or
equivalently, its minimum eigenvalue is positive. In figure
\ref{firstdec} we have represented the minimum of these
eigenvalues as a function of the constant $k$. There exists a
range of $k$ where $P$ is positive, which means that the witness
is decomposable (for instance taking $k=1.2$).

This decomposition gives us insight into the non-local properties
of those states violating the inequality (\ref{tritineq}) with the
optimal settings described above. First, all the states with
positive partial transposition do not violate this inequality.
Moreover, if a given state $\rho$ is detected by the corresponding
witness, $\tr(\rho W)<0$, it follows that $\tr(\rho P_a^{T_A})<0$,
and since $P_a^{T_A}=1/2(\opone-d\¦\Psi\ra\la\Psi\¦)$, we have
\begin{equation}
\label{distcond}
  \la\Psi\¦\rho\¦\Psi\ra>\frac{1}{d},
\end{equation}
which means that the state is distillable \cite{Horo4}. Thus, for
these settings, Bell inequality violation implies state
distillability \cite{AS}.

It would be nice that the recipe for finding this decomposition
was general and that it worked for other choices of the
experimental settings, or higher dimension. Unfortunately this is
not the case. Indeed we have seen numerically that the same
procedure does not work for other settings violating
(\ref{tritineq}), i.e. such that there exists a state that is
detected by the corresponding witness or Bell operator. We have
also considered higher dimensional systems, with the corresponding
Bell inequality (\ref{genineq}) and optimal settings. A similar
decomposition is found for $d=4$ but our method fails for $d=5,6$.
Looking at the variation of the Bell operator spectrum (for the
optimal settings), when $d$ increases, we can somehow understand
why this procedure does not work anymore. For $d=3$, ($d=4$), the
maximum eigenvalue of $B$ is $\simeq 2.9149$ ($\simeq 2.9727$),
while for the maximally entangled state, having the same Schmidt
basis, we have $\simeq 2.8729$ ($\simeq 2.8962$). This may mean
that this maximally entangled state is quite close to the region
of maximal violation, and then (\ref{distcond}) holds for those
states violating the inequality. For higher dimension the
difference between these two values increases (3.0517 vs 2.9105
for $d=5$), what can explain why the decomposition is not
possible. Let us mention here that in the case of two qubits we
are able to find $P$ and $Q$ for any witness associated to the
CHSH inequality following a similar approach \cite{ALP} (actually,
it is a known result that any two-qubit entanglement witness is
decomposable \cite{Horo}). Whether a better generalization of the
CHSH inequality exists satisfying also this construction is an
interesting open question.

\section{Concluding remarks}

In this work we analyzed different manifestations of quantum
non-locality in two-qutrit systems. The starting point was the
Bell inequality introduced in \cite{CGLMP,KKCZO}.

First, we observed that the largest violation of the inequality
over all possible Von Neumann measurements and initial states is
not obtained for the maximally entangled state of two qutrits. We
also proved that the resistance to noise is not a good measure of
non-locality, and we proposed some simple alternatives. However,
even for these measures, the maximal resistance seems not to be
given by the maximally entangled state (further research in this
direction is needed). Our results suggest that two Von Neumann
measurements per site do not detect two-qutrit non-locality in an
optimal way. Notice that the results in table \ref{summary} may
imply  that indeed two measurements on each side become less
efficient for the detection of non-locality when the dimension
increases. It is not excluded that POVMs would give a larger Bell
inequality violation for the two-qutrit maximally entangled state.
However it seems more likely that more observables on each side
are needed. In this sense, we still lack a good generalization of
the CHSH inequality to qutrits.

We also related the violation of this inequality to other
manifestations of non-locality. For the optimal settings, we
demonstrated that the corresponding witness is decomposable by
explicitly deriving its decomposition (\ref{decomp}) in terms of
$P$ and $Q$. Moreover, our construction sheds light into the
distillability properties of those states violating the
inequality. In fact, bound entangled states do not violate the
inequality for the optimal settings (see also \cite{KZG}).

Our last point is more general and concerns the characterization
of non-locality. As it has been mentioned, the resistance to noise
has proven to be an incorrect criterion for the analysis of
non-locality. Is a single number enough for describing all the
non-local features of quantum states for bipartite systems of
dimension greater than qubits, even for the pure-state case? Or,
as it happens for other questions related to entanglement, are
more parameters needed?


\section{Acknowledgements}

We thank Marek Zukowski for many useful comments. A. A. and N. G.
acknowledge financial support from the Swiss FNRS and OFES, within
the European project EQUIP, and the Swiss NCCR ''Quantum
Photonics". A. A. also thanks the Spanish MEC and ESF-QIT for
funding. T. D. is a postdoctoral fellow of the Flemish Fund for
Scientific Research (FWO). J. I. L. acknowledges financial support
from CICYT (AEN99-0766), CIRIT (1999SGR-00097) and EQUIP
(IST-1999-11053).

\section*{ Appendix: maximization of the Bell inequality violation by
entangled qutrits}

Consider the situation that was described in section 2. Two
three-dimensional subsystems $A$ and $B$ are prepared in an
arbitrary entangled state $\vert\Phi\ra\in\sp^3\otimes\sp^3$.
Then, the two parties apply a unitary operation on each subsystem
with only non-zero terms in the diagonal equal to $e^{i\phi_a(j)}$
for $A$ and $e^{i\varphi_b(j)}$ for $B$, with $j=0,1,2$ and
$a,b=1,2$. The setting of the 6 phases $\phi_a(j)$ and
$\varphi_b(\tilde j)$ ($j=0,1,2$) defines an experimental
configuration. Later, $A$ carries out a discrete Fourier
transform, $U_{FT}$, and $B$ applies $U^*_{FT}$. Finally, at the
output of such devices, the states are detected in the local bases
$\vert k_A\ra$ and $\vert l_B\ra$. When the state is prepared
initially in the pure state $\vert \Phi\ra=\Sigma_{j,\,\tilde
j\,=0}^2 \alpha_{j,\,\tilde j}\vert j\tilde j\ra$ the probability
of a joint detection in the $k$th detector in $A$ and in the $l$th
detector in $B$ is equal to
\begin{equation}
P(k,l)={1\over 9} \vert \Sigma_{j,\tilde j=0}^2
e^{i(\phi_a(j)+\varphi_b(\tilde j)+(jk-\tilde jl){2\pi \over
3})}\alpha_{j,\tilde j}\vert ^2.
\end{equation}
It was shown in \cite{Zukowski} by numerical methods that when the
state is maximally entangled (\ref{maxent}), the maximal
resistance of non-locality against noise, when all the possible
phases $\phi_a(j), \varphi_b(j)$ are considered (with $a,b=1,2$
and $j=0,1,2$) corresponds to the optimal phase-settings
(\ref{phase}).

Let us now consider the associated Bell inequality
(\ref{tritineq}). When (\ref{phase}) is satisfied and we let vary
the state $\vert \Phi\ra\in\sp^3\otimes\sp^3$, the violation of
this inequality is maximal when the state $\vert \Phi\rangle  $
is, up to normalization, $\vert 0 0\ra+\gamma\vert 1 1\ra+ \vert 2
2\rangle$, with $\gamma=(\sqrt 11-\sqrt 3)/2\simeq 0.7923$. The
violation is equal to $1+\sqrt{11/ 3}\approx 2.91485$.

Indeed, take the general state $\vert\Phi\ra=\Sigma_{j,\tilde
j=0}^2 \alpha_{j,\tilde j}\vert j\tilde j\ra$ (where
$\Sigma_{j,\tilde j=0}^2 \vert \alpha_{j,\tilde j}\vert^2$ = 1 by
normalization). Then, (\ref{tritineq}) can be rewritten as
follows:
\begin{eqnarray}
\label{i3}
I_3(\vert \Phi\ra)&=&{1\over 9}\Sigma_{j,\tilde
j,m,\tilde m=0}^2\alpha^*_{m,\tilde m}\alpha_{j,\tilde j}
\Sigma_{k=0}^2
e^{i{2\pi\over 3}(k(j-m)-k(\tilde j-\tilde m))} \nonumber\\
&&\left(e^{i(\phi_1(j)-\phi_1(m)+\varphi_1(\tilde
j)-\varphi_1(\tilde
m))}(1-e^{-i{2\pi\over 3}(j-m)}) \right. \nonumber\\
&&+ \left. e^{i(\phi_2(j)-\phi_2(m)+\varphi_1(\tilde
j)-\varphi_1(\tilde m))}(e^{-i{2\pi\over 3}(\tilde j- \tilde
m)}-1) \right. \nonumber\\
&&+ \left. e^{i(\phi_2(j)-\phi_2(m)+\varphi_2(\tilde
j)-\varphi_2(\tilde m))}(1-e^{-i{2\pi\over 3}(j-m)}) \right.
\nonumber\\
&&+ \left. e^{i(\phi_1(j)-\phi_1(m)+\varphi_2(\tilde
j)-\varphi_2(\tilde m))}(1-e^{i{2\pi\over 3}(\tilde j- \tilde
m)})\right) .
\end{eqnarray}
$I_3(\vert \Phi\rangle  )$ can be expressed as $\langle\Phi\vert
B\vert \Phi\ra$ where $B$ is the (self-adjoint) Bell operator. The
maximal value of $I_3$ is thus reached when $\vert \Phi\rangle $
is the eigenstate associated to the maximal eigenvalue of $B$,
$\¦\Psi_{mv}\ra$. We must now determine what this eigenvalue is.
The problem is considerably simplified if we note that
$\Sigma_{k=0}^2 \,e^{i{2\pi\over 3}k(p-q)}=3\,\delta^{(3)}_{pq}$,
where $\delta^{(3)}_{pq}=1$ when $p=q$ modulo 3 and 0 otherwise.
This means that $B$ can be decomposed into the sum of three
operators that are decoupled and act individually inside the
subspaces spanned by the vectors $\{\vert 00\ra,\vert 11\ra,\vert
22\ra\}$, $\{\vert 01\ra,\vert 12\ra,\vert 2 0\ra\}$ and $\{\vert
02\ra,\vert 10\ra,\vert 21\ra\}$ respectively. Inside the subspace
spanned by the vectors $\{\vert 00\ra,\vert 11\ra,\vert 22\ra\}$,
$j-m=\tilde j-\tilde m$ and the reduced Bell operator obeys the
following equation:
\begin{equation}
\left(B^{red1}\right)_{mj}={2\over 3}\{e^{i{2\pi\over3}{(j-m)\over
4}}-e^{i{2\pi\over3}{(-3)(j-m)\over
4}}+e^{i{2\pi\over3}{(-1)(j-m)\over
4}}-e^{i{2\pi\over3}{3(j-m)\over 4}}\},
\end{equation}
and in matricial notation,
\begin{equation}
B^{red1}={2\over 3}\left(\begin{array}{ccc}
0 & \sqrt 3 & 3 \\
\sqrt 3 & 0 & \sqrt 3 \\
3 & \sqrt 3 & 0 \\
\end{array}\right) .
\end{equation}
Inside the two other subspaces, we obtain in a similar fashion the
following expression for the reduced Bell operator:
\begin{equation}B^{red2}=B^{red3}={2\over 3}\left(\begin{array}{ccc}
0 & \sqrt 3 & 0 \\
\sqrt 3 & 0 & 0 \\
0 & 0 & 0 \\
\end{array}\right) .
\end{equation}
The problem consists now of determining the maximal eigenvalues of
these 3x3 matrices. One can check directly that in this matricial
notation $\¦\Psi\ra$ is not an eigenstate of $B^{red1}$ so that it
does not certainly maximize the violation of the Bell inequality.
The eigenvalues of $B^{red1}$ are equal to $-2,1-\sqrt{11/3}$ and
$1+\sqrt{11/3}$, while for $B^{red2,3}$ we have $-2/\sqrt{3}$, 0
and $2/\sqrt{3}$, so that the maximal violation is equal to
$1+\sqrt{11/3}\approx 2.91485$. It is easy to check that the
corresponding eigenvector is, up to normalization, $\vert 0
0\ra+\gamma\vert 1 1\ra+ \vert 2 2\rangle$, with $\gamma=(\sqrt
11-\sqrt 3)/2\simeq 0.7923$.

Note that at first sight it could seem strange that the maximally
entangled state does not maximize the violation of the inequality
(\ref{tritineq}), because it seems that the states $\vert 00\ra$,
$\vert 11\ra$ and $\vert 22\ra$ are undistinguishable in our
approach. The discrete Fourier transforms are well-known for their
cyclic properties, and in this inequality all the detectors are
treated on an equal foot (the Bell operator contains cyclic
summations of probabilities of coincident firings). Nevertheless
if we consider (\ref{i3}), we can notice that the matrix
coefficient $\langle m\tilde m\vert B \vert j\tilde j\rangle$
contains expressions of the type
$e^{i(\phi_a(j)-\phi_a(m)+\varphi_b(\tilde j)-\varphi_b(\tilde
m))}$ where $\phi_a(j)$ and $\varphi_b(\tilde j)$ are locally
adjustable phases. Due to the presence of factors of this type,
the cyclic invariance is broken (in the sense that when
$j-m=j'-m'$ modulo 3 and $\tilde j-\tilde m=\tilde j'-\tilde m'$
modulo 3, it is not necessarily true that $\langle m\tilde m\vert
B \vert j\tilde j\ra=\langle m'\tilde m'\vert B \vert j'\tilde
j'\rangle$). Note that when the phase settings are optimal, they
depend linearly on the indices $j,m,\tilde j,\tilde m$ according
to (\ref{phase}) and $\la m\tilde m\vert B \vert j\vert \tilde
j\rangle$ depends on $j,m,\tilde j,\tilde m$ only through the
combinations $j-m$ and $\tilde j -\tilde m$. Thus, the matrix
coefficients $\langle m\tilde m\vert B \vert  j\tilde j\rangle$
and $\langle (m + \delta)(\tilde m+ \tilde \delta)\vert B \vert (j
+ \delta)(\tilde j+\tilde \delta)\rangle$ are equal. This explains
why $B^{red2}=B^{red3}$ and also why the states $\vert 00\rangle $
and $\vert 22\rangle  $ appear symmetrically in the matrix
$B^{red1}$. Nevertheless, the cyclic invariance is still broken
(in the sense made precise above) which explains why $B^{red1}$ is
singularized relatively to $B^{red2}$ and $B^{red3}$, as $\vert
11\rangle  $ relatively to $\vert 0 0\rangle  $ and $\vert 2
2\rangle  $.

Finally, we prove that we must not expect a larger violation of
the inequality (\ref{tritineq}) if we modify the phase settings
defined in (\ref{phase}). Indeed, let us vary $\vec\phi_1$,
keeping the other phases fixed:
\begin{equation}
\label{phasevaried} \phi_1(0)=0 \quad\quad \phi_1(1)=\alpha
\quad\quad \phi _1(2)=\beta \quad\quad \phi_2(j)={\pi\over 3} j
\quad\quad \varphi_1(j)={\pi\over 6}j \quad\quad
\varphi_2(j)=-{\pi\over 6}j.
\end{equation}
Then,
\begin{equation}
\label{Ovar} B_{varied}^{red1}={1\over
3}\left(\begin{array}{ccc}  0 & \sqrt 3 & 3 \\
\sqrt 3 & 0 & \sqrt 3 \\
3 & \sqrt 3 & 0 \\
\end{array}\right)+{1\over 3}\left(\begin{array}{ccc}
0 & \sqrt 3 e^{i\alpha}& 3 e^{i\beta}\\
\sqrt 3  e^{-i\alpha}& 0 & \sqrt 3 e^{i(\beta-\alpha)} \\
3 e^{-i\beta} & \sqrt 3 e^{i(\alpha-\beta)} & 0 \\
\end{array}\right) .
\end{equation}
It is easy to check that the two matrices that appear in
(\ref{Ovar}) have the same eigenvalue equation and thus the same
spectrum, which is $-1$, $(1-\sqrt{11/3})/2$ and
$(1+\sqrt{11/3})/2$. The maximal eigenvalue of these matrices is
also equal to their norm in the present case, so that the norm of
$B_{varied}^{red1}$, and thus its maximal eigenvalue, is certainly
not larger than $1+\sqrt{11/3}$. Entirely similar results can be
obtained when $\vec\phi_2$, $\vec\varphi_1$ or $\vec\varphi_2$ are
varied. This proves that the phase settings defined in
(\ref{phase}) maximize locally (in the space of all possible
phase-settings) the violation of the Bell inequality
(\ref{tritineq}).

\newpage

\begin{table}
\begin{tabular}{||c|c|c|c||}
  ~~~~~~Dimension~~~~~~ & Violation for $\¦\Psi\ra$~~~~~~~~ & Maximal violation (for $\¦\Psi_{mv}\ra$)~~~~~~~~ & Difference ($\%$)~~~~~~~~\\ 
\hline 3 & 2.8729~~~~~~ & 2.9149~~~~~~ & 1.4591~~~~~~ \\
\hline 4 & 2.8962~~~~~~ & 2.9727~~~~~~ & 2.6398~~~~~~ \\
\hline 5 & 2.9105~~~~~~ & 3.0157~~~~~~ & 3.6133~~~~~~ \\
\hline 6 & 2.9202~~~~~~ & 3.0497~~~~~~ & 4.4345~~~~~~ \\
\hline 7 & 2.9272~~~~~~ & 3.0776~~~~~~ & 5.1411~~~~~~ \\
\hline 8 & 2.9324~~~~~~ & 3.1013~~~~~~ & 5.7588~~~~~~ \\
\end{tabular}
\medskip
\caption{Violation of the inequality (\ref{genineq}) for two
qudits, $\sp^d\otimes\sp^d$, up to $d=8$. It is shown the value
obtained for the maximally entangled state (\ref{maxent}) and the
maximal violation of the inequality corresponding to the largest
eigenvalue of the Bell operator.} \label{summary}
\end{table}


\begin{references}
\bibitem{sep}
Take a composite system of $N$ particles,
$\sp^{d_1}\otimes\sp^{d_2}\otimes\cdots\otimes\sp^{d_N}$, where
$d_i$ is the dimension of the local space of party $i$. A pure
state, $|\Phi\ra$, belonging to the global space, is said to be
separable when it can be written as a tensor product of pure
states in each subsystem, i.e.
$|\Phi\ra=|\psi_1\ra\otimes|\psi_2\ra\otimes\cdots\otimes|\psi_N\ra$.
This definition is easily extended to mixed states, and a density
matrix $\rho$ is separable when it can be expressed as a convex
sum of projectors into separable pure states, i.e. $\rho=\sum_i
p_i |\Psi^s\ra\la\Psi^s|$, where $|\Psi^s\ra$ are separable.
Physically, separable states are those that the different parties
can prepare if they are allowed to use only local operations in
each local system and to communicate classically. States that are
not separable are called entangled.
\bibitem{fourier}
The matrix elements of the Fourier transform in $\sp^d$ are given
by $\left(U_{FT}\right)_{jk}=e^{i(j-1)(k-1)2\pi/d}$.
\bibitem{note}
Note that for the case in which the entangled state is pure, the
expression for the closest separable state is known \cite{relent}.
\end{references}
\end{document}